# On an Atom-Interferometer Test of the Speed of Collapse of Wavefunctions in Relativistic Quantum Mechanics

**Hans C. Ohanian**

(hohanian@uvm.edu)

*Department of Physics*
*University of Vermont, Burlington, VT 05405-0125*

**Abstract.** I propose to resolve the controversy over the speed of collapse of quantum-mechanical wavefunctions by means of an experimental test with a modified symmetric Mach-Zehnder atom interferometer, with non-intersecting, parallel, widely separated final beams. According to the conventional collapse scenario, the coherent twin-peak atomic wavefunction in the beams of the interferometer suffers an instantaneous collapse, at infinite speed, when the atom is captured by one of the two detectors at the ends of the beams, but it remains coherent until that instant. In contrast, according to the Hellwig-Kraus relativistic collapse scenario, the wavefunction collapses at the speed of light, backward in time along the past light cone of each detector. This leads to a premature collapse, or pre-collapse, which for a beam-to-beam separation of 3 m extends over a time span of 10 ns before arrival at the detectors. Within this time span the paired wavepackets in the two beams will be incoherent. The difference between the coherent wavepackets of the conventional scenario and the incoherent wavepackets of the relativistic scenario can be tested by probing the atomic beams with a transverse laser beam crossing them near the detectors. If the paired atomic wavepackets in the two beams are coherent, the light scattered by the two beams will also be coherent and generate a standing light wave in the space between the beams, with detectable interference fringes. If the paired wavepackets are incoherent, no such interference fringes will be generated, and the distribution of the scattered light will be that of two independent dipoles. The design parameters for this test seem to lie within reach of the techniques of atom interferometry.

## I. INTRODUCTION

Of the many mysteries that shroud the foundations and the interpretation of quantum mechanics none has been as much neglected as the mystery of the relativistic propagation of the collapse of wavefunctions during measurements.[1] In nonrelativistic quantum mechanics this collapse is usually supposed to occur instantaneously, at infinite speed along a constant-time hypersurface of an inertial reference frame. But in relativistic quantum mechanics such an instantaneous collapse is nonsensical, because an instantaneous spacelike interval in one inertial reference frame is not necessarily instantaneous in another. Expressed otherwise, what is an infinite speed of collapse in one reference frame becomes a finite speed in another, as is obvious from the Lorentz transformation for velocity, which tells us that an infinite signal velocity $v \to \infty$ becomes finite (and possibly negative!) in a new reference frame of relative velocity $V$, that is, $v' = (v-V)/(1-vV/c^2) \to -c^2/V$. Such discrepancies in propagation of collapse in different reference frames lead to violations of probability-conservation laws (violations of unitarity), and violations of other conservation laws, and also failure of the generally accepted Lorentz-transformation laws for probability as an invariant scalar density and failure of the corresponding transformation laws for relativistic wavefunctions.[2] This has dire consequences for Wigner's conception of quantum-mechanical wavefunctions as representations of the Lorentz group (scalars, vectors, spinors, etc.), and it undermines our reliance on Lorentz symmetry in the construction of invariant Lagrangians and conservation laws for the fundamental interactions of high-energy physics.

Despite the disastrous consequences lurking in the relativistic collapse of wavefunctions, the theoretical and experimental exploration of these problems has received insufficient attention.[3] Even in the writings on quantum mechanics by some of our best theorists these problem are often completely ignored. For instance, Weinberg's recent textbook *Lectures on Quantum Mechanics* [4] examines several interpretations of quantum mechanics in commendable detail, but says nothing whatsoever about the relativistic collapse problems that infest these interpretations. And such discussions as are available in the literature often veer into the Land of Oz. In one of their publications, Aharonov and Albert [5] suggested that in view of the failure of probability conservation arising from the Lorentz transformation of instantaneous collapse we must contemplate that one particle might sometimes transmogrify into two or into none (with an ensuing failure of conservation of energy, charge, baryon number, etc.?).

Most of the attempts to repair the problems of relativistic collapse rely on the rather naïve and arbitrary stipulation that the speed of collapse is infinite in a selected preferential reference frame, and that wavefunctions and probabilities before and after collapse are to be calculated exclusively in this reference frame. Only after this calculation in the preferential reference frame are these quantities to be transformed to other reference frames. Direct calculations in such other reference frames are to be avoided, and any inconsistencies in the probability-conservation laws and concomitant conservation laws are to be cheerfully ignored ("See no evil, hear no evil, speak no evil"). This nonchalant attitude rests on the lame excuse that these inconsistencies are only intermittent, restricted to naughty patches of spacetime near measurements.[6]

Much of this is rather reminiscent of the status of electrodynamics in the pre-relativistic era, when it was presupposed that Maxwell's equations were valid only in a preferential ether frame and that the use of the same equations in any other reference frames was improper. This era came to an end when the Michelson-Morley interferometer experiment gave us decisive evidence for an invariant speed of light and a new relativistic formulation of physics.[7] By analogy, it appears desirable to contrive an atom-interferometer experiment on the speed of collapse of quantum-mechanical wavefunctions and obtain decisive evidence for a finite speed of this collapse. The experiment proposed here is intended to find evidence for a collapse at the speed of light, instead of the conventional collapse at infinite speed.

## II. HELLWIG-KRAUS COLLAPSE

The only well-motivated and consistent theoretical formulation of relativistic collapse seems to be that of Schlieder [8] and of Hellwig and Kraus (H-K),[9] published in 1968 and 1970, respectively, but only rarely mentioned in the voluminous literature on the foundations of quantum mechanics.[10] Instead of instantaneous collapse of the quantum-mechanical wavefunction along a constant-time hypersurface, Schlieder and Hellwig-Kraus proposed collapse along the past light cone of the spacetime point at which the measurement is performed, what might be called collapse backward in time at the speed of light. The mathematical treatment of this collapse is most conveniently handled in the Heisenberg representation, with time-independent wavefunctions, but time-dependent operators.[11] In this representation, the collapse triggered by a measurement at a given spacetime point produces a new time-independent wavefunction that engulfs all of spacetime *except* the interior of the past light cone (see Fig. 1). Expressed in topological jargon, the new, collapsed, wavefunction resulting from the measurement extends over the complement of the interior of the past light cone.

In the Schrödinger representation the new wavefunction can be completely characterized by its boundary conditions along the surface of the past light cone, provided the system has a finite spatial extent and the wavefunction is zero in the surrounding empty space. In this representation, we can therefore say that the new wavefunction extends over the surface of the past light cone, and its time evolution is to be calculated from these given initial boundary conditions by means of the appropriate relativistic wave equations (but these boundary conditions are not applicable to calculations that extend into the interior of the past light cone).

The H-K scenario has various advantages over instantaneous collapse along a constant-time hypersurface, and it eliminates all the several problems associated with relativistic collapse listed in Section I. In essence, the H-K scenario is merely a kinematical algorithm that makes the collapse scenario consistent with the geometry of relativistic spacetime and Lorentz symmetry. It does not provide any dynamical foundation for wavefunction collapse, which is to be regarded as a separate though not entirely independent issue—any proposal for a relativistic dynamical mechanism of wavefunction collapse that does not obey the relativistic kinematics of the H-K scenario falls under vehement suspicion of a logical flaw.

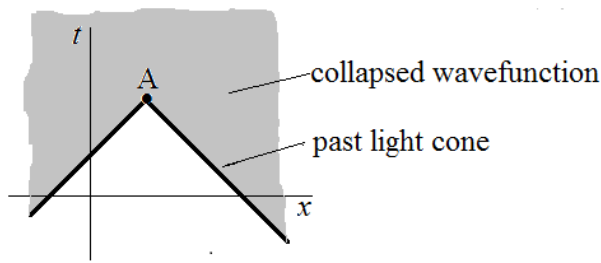

Fig. 1

FIG. 1 Hellwig-Kraus relativistic collapse of a wavefunction triggered by a measurement at the spacetime point A. In the Schrödinger representation the new collapsed wavefunction extends over all of spacetime (gray region), except the interior of the past light cone (white region). The new wavefunction is undefined (blank) in the interior of the past light cone. [This dichotomy is to be expected, because the information gained by the measurement requires a revision of the probabilistic predictions for all regions that, relative to the point A, are in the future or might be in the future in any reference frame $x',t'$ obtained from the reference frame $x,t$ by a Lorentz transformation; in contrast, in the interior of the past light cone—that is, in the absolute past—the new available information invalidates the old probabilities calculated for this region and leaves them undefined.] Each additional successive measurement triggers another similar collapse at the next such past light cone.

Note that in the limit of non-relativistic quantum mechanics, for systems of particles moving with low speeds, the H-K scenario reduces to conventional collapse along a constant-time hypersurface.

Several objections have been raised against the H-K scenario,[12, 13] but I believe that these objections have little merit. Mould has presented a solid defense against these objections,[14] and I hope to review the arguments and counterarguments in a later paper. For now, I do not wish to become entangled in a contentious debate about these objections. Instead, I propose to resolve the controversy over H-K collapse by a straightforward experiment, which, I believe, is almost within the reach of current techniques used in atom interferometry.

## III. PRE-COLLAPSE

The proposed experiment relies on what is the weirdest aspect of H-K collapse, namely, that under some circumstances, the H-K scenario produces a pre-collapse, that is, the wavefunction collapses before the measurement is performed. A simple example of such a pre-collapse arises when we consider an atom whose wavefunction is equally distributed over the two beams of an atom interferometer. This is a "schizoid" atom, about which we cannot say whether it is in the left beam or the right beam—we can say only that there is a 50-50 chance that we will find it left or right when we place detectors at the ends of the beams and perform measurements that reveal the presence or absence of the atom. Before such a "which-beam" measurement (sometimes called a *welcher Weg* measurement), the atom is described by a coherent twin-peak wavefunction, consisting of a superposition of one sharply concentrated wavepacket in the left beam and another sharply concentrated wavepacket in the right beam. If the detector at the left beam captures the atom, the wavefunction collapses to a single wavepacket on the left; if the detector at the right beam captures the atom, the wavefunction collapses to a single wavepacket on the right. Obviously, capture by a detector is a measurement. But, in the manner of EPR, non-capture by a detector is also a measurement—if the detector does *not* capture the atom (and is 100% efficient), then this non-capture allows us to conclude that the atom has been captured by the *other* detector, and we can regard the measurement completed, even without a direct check of the other detector.[15]

In the H-K scenario, with 100% efficient detectors, such a two-detector arrangement actually produces two H-K collapses: one collapse along the past light cone of the spacetime point A where the left detector operates and one collapse along the past light cone of the spacetime point B where the right-hand detector operates (see Fig. 2). If A and B have a spacelike separation, these two distinct collapses are justified by the causal independence of the measurement results at A and B—each of these spacetime points is outside of the future light cone of the other, so each can trigger a causally distinct H-K collapse of its own (subject to the constraint of the probability-conservation law that demands opposite capture vs. non-capture results at A and B).[16] The only region of spacetime in which the wavefunction does not collapse is the region of overlap of the absolute pasts of A and B, or in topological jargon, the region of intersection of these absolute pasts (shown white in Fig. 2).

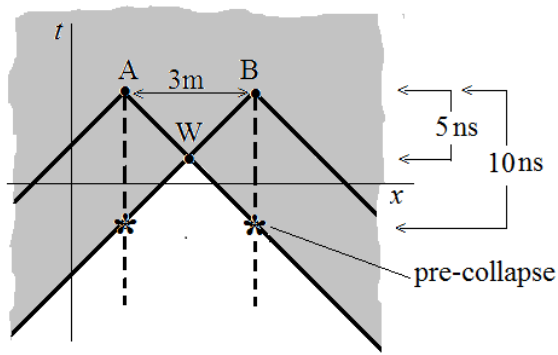

Fig. 2

FIG. 2 Pre-collapse of a wavefunction triggered by joint measurements at two spacetime spacetime points A and B, assumed to be simultaneous (these spacetime points could be any two points with a spacelike separation; here we take them to be simultaneous for the sake of convenience). The wavefunction collapses in all of spacetime except in the white region, which is the topological intersection of the absolute pasts of A and B. In the diagram shown here, with only one space dimension, this white region looks like a cone, but in three space dimensions the surface of the white region has a more complicated shape and lacks rotational symmetry about its central line. The apex of the white region occurs at a point W, which in the $x,t$ -reference frame is earlier than the points A and B. For illustrative purposes, this diagram assumes a spatial separation of 3 m between the points A and B. If the wavefunction extends over all the space between A and B, the pre-collapse time is 5 ns. But if the wavefunction is concentrated in two sharp peaks separated by 3 m proceeding along approximately parallel worldlines (like the dashed lines shown here, which represent the worldlines of the twin peaks of the coherent wavefunction of an atom in the two separate beams of an interferometer), the pre-collapse time is almost 10 ns.

As we can see from Fig. 2, the joint collapse extends to an earlier time than that of the spacetime points A and B, that is, it involves a premature collapse, or pre-collapse. For instance, if the transverse distance between A and B is 3 m and if the wavefunction extends over all the space between A and B, then the pre-collapse in the $x$, $t$- reference frame extends back in time by 5 ns (which corresponds to the spacetime point W). If the wavefunction is concentrated in two sharp peaks to the left and the right of W, then its backward extrapolation can avoid entering the white region for more than 5 ns, and, correspondingly, the pre-collapse occurs earlier than W. Thus, the sharply concentrated twin wavepackets of an atom proceeding coherently along the parallel beams of an atom interferometer with a 3-m beam-to-beam separation attains pre-collapse at a time of about 10 ns before the detection of the wavepackets at A and B.[17]

This pre-collapse is an instance of the "spooky action-at-a-distance" of the EPR experiment, now made even spookier by action into the past. The pre-collapse at the left boundary of the white region in Fig. 2 is triggered by the measurement at the right-hand detector (B), and vice versa. The measurement at the left detector (A) cannot pre-collapse the wavefunction along the left beam, because such a pre-collapse into the absolute past of A would be a violation of causality. But for the right-hand detector, the pre-collapse of the wavepacket along the left beam is not into the absolute past, and the causality problem does not arise.

## IV. AN EXPERIMENTAL TEST OF HELLWIG-KRAUS COLLAPSE

The pre-collapse phenomenon offers a good opportunity for an experimental test of the H-K scenario. Figure 3 is a rough sketch of the proposed experimental arrangement. The apparatus is a modification of the Mach-Zehnder atom interferometer used by Chapman et al.[18] in their ingenious experiment that investigated how the coherence of the wavefunction in the beams of the interferometer is degraded by illumination with photons and how the beams lose their ability to interfere when photon scattering provides information about the path of the atom.

As in the original experiment of Chapman et al., a beam of sodium atoms enters the interferometer, and the first diffraction grating (at the bottom of Fig. 3) splits the beam into two spatially separated coherent beams. In the original experiment, the grating had a spacing of 200 $\mu$m and the two beams were separated by only a couple of mm; for our purposes, a smaller grating spacing and a consequently larger beam separation would be better. The next two gratings shown in Fig. 3, or maybe several more pairs of such gratings, increase the separation between the beams and finally make them approximately parallel.[19] However, in contrast to the Mach-Zehnder interferometer, the final parallel beams in Fig. 3 are not reunited to examine their interference. Instead, the atoms in the beams are detected by a pair of detectors (such as hot-wire detectors) at the ends of the beams.

The detection of an atom in one detector and the nondetection in the other constitute two measurements at A and B, and, as in Fig. 2, the H-K scenario implies a pre-collapse of the twin wavepackets.[20] We can discriminate experimentally between this pre-collapse and the conventional instantaneous collapse by borrowing another feature from the Chapman et al. experiment, namely, a transverse laser beam that crosses the atomic beams and provokes fluorescent scattering of its light by the atomic wavefunction. For our test, we place the transverse laser beam very close to the ends of the two atomic beams, where the hot-wire detectors measure the atom position and collapse the wavepackets.

If these final position measurements trigger pre-collapses along the past light cones of the detection points A and B, the twin wavepackets in the atomic beams will become decoherent at a location below A and B, but slightly above the intersection of the atomic beams with the laser beam, and the fluorescent radiation emitted by the twin wavepackets will then be that of two incoherent dipole antennas.[21] In contrast, if the final position measurements trigger a conventional collapse, instantaneously at the simultaneous detection points A and B, the twin wavepackets will remain coherent until they reach these points, and the radiation pattern will be that of two coherent dipole antennas, with a phase difference $2\pi d / \lambda$, where $d$ is the distance between the beams and $\lambda$ the wavelength of the laser light. In the rest frame of the moving wavepackets, the radiation in region between the wavepackets then consists of a standing e.m. wave, with interference fringes separated by a transverse distance of $\lambda/2$.[22]

After the atom wavepackets reach the detection points A or B, any further radiation emitted is always incoherent and therefore does not help to discriminate between H-K collapse and conventional collapse. Besides, the atomic states will probably be mangled by the collisional interactions with the atoms of the detectors, and there might not be any such further fluorescent radiation at all.

If the radiation emitted before the detection points is coherent and produces a standing wave e.m. wave in the rest frame of the atomic wavepackets, this can be detected in the laboratory frame by placing a narrow strip of photographic film along the midline between the two atomic beams, so the standing-wave maxima sweeping longitudinally through this film imprint a pattern of closely-spaced parallel longitudinal fringes in the photographic emulsion (the capture of fringes on the scale of a wavelength or less requires a fine-grained photographic emulsion, such as used for the production of holograms). Each pair of wavepackets for each atom that passes through the interferometer contributes the same intensity pattern of fluorescent light waves at the location of the film strip, so adequate statistical data to prove or disprove the existence of interference fringes can be accumulated by passing a large number of atoms through the interferometer. If the experiment reveals interference fringes, it would refute the H-K collapse scenario.[23]

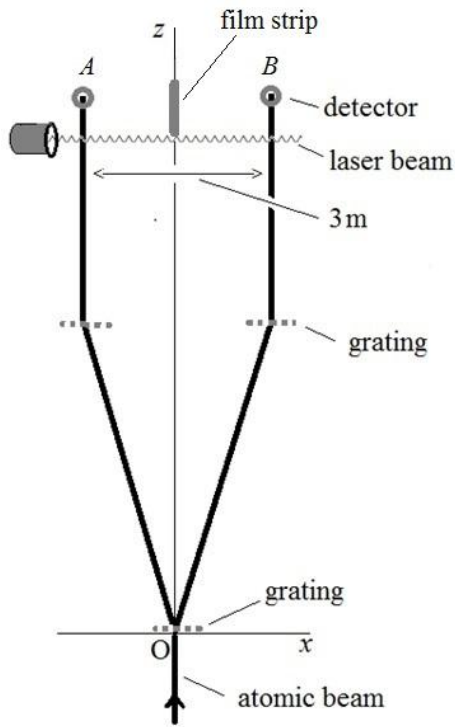

Fig. 3

FIG. 3 A modified version of the Mach-Zehnder atom interferometer of Chapman et al. The atomic beam enters the interferometer at O, where a grating splits the beam into two coherent beams of equal amplitudes. Two other gratings then make these beams approximately parallel and aim them at the detectors A and B. Near the detectors, a laser light beam crosses the two atomic beams and provokes fluorescent emission of scattered light. Some of this light is captured by the film strip at the center.

In the Chapman et al. experiment the intensity of the laser was adjusted to achieve maximum excitation of the atom within the time needed to cross the focal spots of the laser beam at the intersections with the atom beams. Furthermore, the atoms had available sufficient time after that crossing, so most of them were able to complete the spontaneous decay (the lifetime of the excited sodium state is 16 ns [24]). If we allow similar times in our modified experiment, we ensure the emission of fluorescent radiation.

Figure 4 summarizes the "time-line" of the critical events during the final moments before impact of an atom on the detectors. The numbers in this figure are relevant distances and times for an assumed beam speed of 3000 m/s. Each atom that passes through the interferometer has such a worldline diagram. The net intensity of the emitted fluorescent light is proportional to the net atom current passing through the interferometer.

The experimental test described here is intended as a proposal for an actual experiment, not merely a hopelessly impractical Gedankenexperiment. To compensate for my deficiencies in the art of experimentation, I deliberately imitated a known successful experiment as closely as possible. But the design still needs elaboration. For instance, we need to overcome the pesky technical problem of the short time available for completing the scattering process. The time spent by a 3000-m/s sodium atom in passing through the 15-$\mu$m focal spot of a laser beam is about 5 ns and the lifetime for the subsequent spontaneous decay is about 16 ns, so the sum of these times exceeds the 10 ns pre-collapse time for a beam-to-beam distance of 3 m. This means we must operate with a reduced available decay time, that is, less than 100% decay. To fix this problem, we might try to decrease the size of the focal spot and supply more laser power.

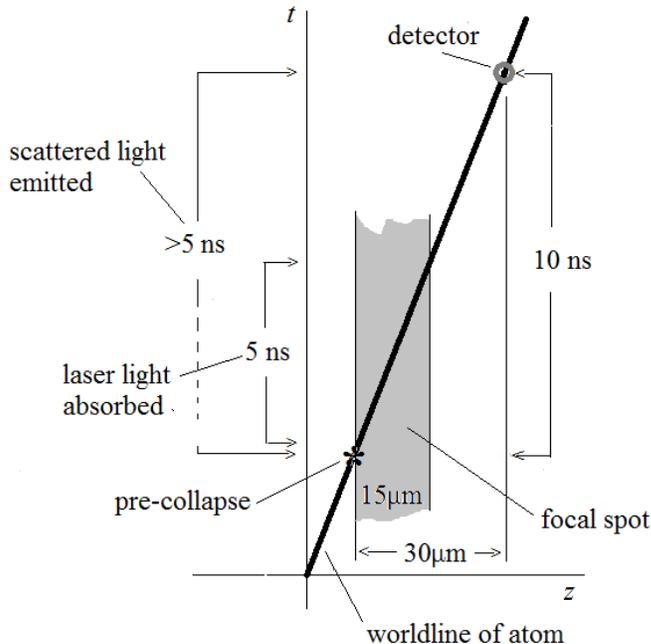

Fig. 4

FIG. 4 Worldlines of the twin wavepackets of an atom traveling through the interferometer. The $z$-axis represents the longitudinal distance along the beam; the scale of this axis has been enlarged, for convenience. The $x$-axis, which represents the transverse distance, has been omitted, so one worldline is hidden behind the other and the two worldlines of the two wavepackets appear as one. The wavepackets take 5 ns to cross the focal spots of the laser beam. The time interval available for spontaneous decay of the excited state partially overlaps the time interval spent in the focal spot; therefore the typical time available for decay is somewhat longer than the 5 ns that remain after crossing the focal spot.

Order-of-magnitude estimates suggest that a beam-to-beam separation of 3 m might be adequate for our purposes. In the Chapman et al. experiment the transverse beam-to-beam separation was only a couple of millimeters. Can we increase this separation by a factor of 1000, or maybe somewhat more, so we can also allow a more generous separation between the laser beam and the hot-wire detectors? There are several ways to achieve larger angular deflections and larger beam-to-beam separations: use gratings with smaller spacings (according to recent reports, improvements in the nano-fabrication of gratings have resulted in the fabrication of gratings with a spacing of 20 nm); use higher orders of diffraction; use several gratings in series; use beams of atoms of lower mass (for a given speed, the de Broglie wavelength of a hydrogen atom is about 20 times as large as that of a sodium atom, and the deflection by a grating is proportionately larger); use an interferometer with longer beams. By judicious combination of such techniques, a beam-to-beam separation of the order of several meters might be attainable.[25]

Another technical problem is the certainty of detection of an atom at A and B. We need to be certain that the atom triggers the which-beam detectors at A or B, but we do not need to gather explicit data about the detection result and actual arrival time. We merely need assurance that the atom wavepackets come into contact with the hot wire or hot ribbon of the detector. Such an intimate contact with a thermal reservoir presumably permits us to affirm that the which-beam measurement is "closed by an irreversible amplification," as demanded by Bohr,[26] and this compels the collapse of the wavefunction. Because the detailed data about the detection are of no direct relevance for our test of H-K collapse, we might not even need a hot-wire detector and its electronic ancillaries—it might be sufficient to provide any kind of penetrable surface (a layer of photographic emulsion or of fly paper?) in which an incident wavepacket can suffer a few collisions with resident atoms. These resident atoms constitute a more or less chaotic environment with statistically random positions and velocities, so during collisions the kinetic energy of the wavepacket changes by random amounts, which leads to random phase deviations of the wavepacket between one collision and the next. A simple estimate shows that for sodium atoms incident on an ordinary solid target such random phase deviations are of the order of magnitude of more than 10 radians per collision, and after just a few collisions the phase of the wavepacket becomes totally random—the statistical randomness of the environment infects the phase. The wavepacket can be regarded as collapsed when total phase randomnization is attained. This process takes only about 0.001 ns.[27]

On the scale of Fig. 4, the new collapsed wavepackets in the two detectors are therefore almost pointlike, like the original uncollapsed wavepackets. The main difference is that along the 10-ns pre-collapsed final segment of the worldline the left and right new wavepackets are incoherent. They also differ slightly in direction, because of the amount of transverse momentum they can acquire during collisions in the detector. On a worldline diagram with equal units along the axes, this transverse momentum is negligible—if $v \ll c$, the four-momentum in such a worldline diagram has a magnitude of $mc$ in the timelike direction, but only $mv_x$ or $mv_z$ in the spacelike directions.

However, when extrapolated backward in time, any transverse momentum $mv_x$ results in a change of position and a corresponding change in the transverse beam-to-beam distance, which affects the phase difference between the fluorescent light contributed by the left and right beams. Fortunately, in the H-K scenario, such an extra random phase difference does no harm. It merely leads to some extra incoherence in the contributions from the left vs. right wavepackets—and this extra incoherence in the fluorescent light does not alter the outcome of the experiment.

It would also be of some interest to explore the radiation pattern of the fluorescent light in full detail. If the H-K scenario is valid, the photons scattered in any selected direction should display (cumulatively) the radiation pattern of two incoherent dipoles, one located at the left beam, one at the right. If instead the conventional instantaneous collapse scenario is valid, the scattered photons should display the radiation pattern of two coherent dipoles.

If the test confirms the H-K collapse scenario we could try to vary various apparatus parameters to discover the physical conditions that are necessary and sufficient to trigger measurement and collapse. Does an "observer" really play any role in this?

Finally, a very brief and oversimplified historical commentary: The experimental investigations of collapse began in the early 1970s with optical EPR experiments by Clauser and Freedman and others on pairs of photons with correlated, entangled, polarizations. These first experiments exploited Bell's theorem to confirm that the wavefunctions of quantum mechanics are entangled over wide regions of space, and they participate in a spooky action-at-a-distance, inconsistent with hidden variables, which allows their distant parts to collapse without any direct physical contact ("nonlocality"). In the early 1980s, Aspect and his collaborators established that the speed of this collapse exceeds the speed of light ("nonseparability"). And in the 1990s, with an experiment on a much larger scale (10 km, at CERN, where monumental experiments are a fad), Gisin and his collaborators established that the speed of collapse exceeds $10^7$ times the speed of light. The proposed atom-interferometer experiment could be said to take this race-to-the-top to a range of speeds beyond infinity, where the collapse proceeds so fast that it travels backward in time.[28]

## APOLOGY AND ACKNOWLEDGEMENT

The original treatment of collapse by Hellwig and Kraus was formulated in the language of quantum field theory. My treatment in Sections II and III is an attempt to translate the field-theoretic treatment into the simple language of particle quantum mechanics and provide a clearer intuitive picture. I apologize for any errors that might have corrupted my translation (as the Italians say, *traduttori, traditori*).

I am grateful to Prof. J. Auping-Birch, SJ, for encouraging me to probe the muddled status of measurement problems in quantum mechanics.

And I thank my friends Tasha and Nekko Bakenekko for their gentle contributions to my thinking.

## REFERENCES AND ENDNOTES

[1] The question of a physical dynamical mechanism for producing the collapse has received considerably more attention, but only in a nonrelativistic context that avoids any examination of how the collapse propagates relativistically through spacetime. R. Penrose, *The Road to Reality*, Knopf, New York (2005), Sections 30.11- 30.13, mentions several proposals for collapse mechanisms, but shies away from a relativistic treatment. In my opinion, any nonrelativistic collapse mechanism is worthless, unless it is based on a relativistic mechanism and can be shown to be a nonrelativistic approximation of the latter. Ditto for interpretations of quantum mechanics and measurements.

[2] See, e.g., I. Bloch, Phys. Rev. **156**, 1377 (1967), Section III. Note that in Section I Bloch is obviously wrong in insisting that the apparatus states must be orthogonal—they must be macroscopically *distinguishable* (that is, they must avoid excessive overlaps), but they need not be exactly orthogonal.

[3] For the sake of full disclosure I have to admit that my own textbook on quantum mechanics [H. C. Ohanian, *Principles of Quantum Mechanics,* Prentice-Hall, Englewood Cliffs (1990)] also suffers from this attention-deficit disorder.

[4] S. Weinberg, *Lectures on Quantum Mechanics,* Cambridge University Press, Cambridge (2013).

[5] Y. Aharonov and D. Z. Albert, Phys. Rev. D **29**, 228 (1984).

[6] See, e.g., Bloch, op. cit., and Aharonov and Albert, op. cit.

[7] Whether Einstein's thinking about relativity was substantially influenced by the Michelson-Morley experiment is unclear. Everybody else's thinking certainly was. See H. C. Ohanian, *Einstein's Mistakes*, W. W. Norton & Co., New York (2008), p. 24.

[8] S. Schlieder, Commun. Math. Phys. **7**, 305 (1968). A suggestion of collapse along the light cone was mentioned even earlier by Bloch (op. cit.), but immediately rejected by him.

[9] K.-E. Hellwig and K. Kraus, Phys. Rev. D **1**, 566 (1970).

[10] The most recent 2004 update of A. Cabello's monumental compendium of 10,000+ references on the foundations of quantum mechanics and the interpretation of measurements (*Bibliographic guide to the foundations of quantum mechanics,* arXiv:quant-phys/0012089v12) lists only about twenty references in its specialized topical section on relativistic collapse, Hellwig-Kraus not among them (although this reference is included in the main, alphabetical, section).

---

[11] I prefer to speak of collapse of the wavefunction rather than the more abstract state vector, because the focus this discussion is the propagation of the collapse in space and time, which requires a spacetime representation of the state vector.

[12] Y. Aharonov and D. Z. Albert, Phys. Rev. D **21**, 3316 (1980); Phys. Rev. D **24**, 359 (1981), Phys. Rev. D **29**, 228 (1984).

[13] O. Cohen, and B. J. Hiley, Foundations of Physics **25**, 1669 (1995).

[14] R. A. Mould, "A defense of Hellwig–Kraus reductions," arXiv: quant-ph/9912044v1 (1999). I concur with almost all of Mould's defense of H-K collapse against the objections listed in Refs.12 and 13. But I think the defense could have been strengthened by emphasizing state reductions by physical means, instead of state reductions by the "distinctive role of the observer." Penrose, op. cit., calls state reductions by purely physical means "objective reductions (**OR**)," and I think he is right in emphasizing objective materialism over subjective idealism (although I am not in favor of Penrose's proposal of objective reduction by gravitational interaction).

[15] To exploit such a non-detection scheme we need to restrict the beam current to a low value, so no more than one atom arrives at the detector within the time interval needed to complete the detection. This sets a limit of about $10^{12}$ atoms/s on the beam current, if we accept an order-of-magnitude estimate of 0.001 ns for the detection time (see endnote 27). In practice, this limit is well satisfied in atom interferometry (typical beam currents reported in the literature are in the range of $10^7$/s to $10^8$/s). For the non-detection scheme we might also need to impose limits on the length of individual atom wavepackets. No great harm is done if the non-detection scheme and the ensuing double collapse were to fail occasionally—this would merely deprive one occasional atom of the opportunity for scattering because of a spoiled pre-collapse.

[16] For instance, if we detect the atom at A, then the second measurement at B that does *not* detect the atom at B can be reckoned as establishing independently that the atom is at A. This would be quite obvious if we knew the expected time of arrival of the atom at B and performed the measurement at that time. But it remains true even if we have no such arrival information, because we can operate the detector at B continuously, so the instant of expected arrival is necessarily one of the instants of operation, and non-detection of the atom at that instant establishes that the atom is located at A, which triggers H-K collapse along the past light cone of the measurement point B. The experimenter is ignorant about exactly which of the non-detections in the time-sequence of detector data at B is an "informative" non-detection that occurs at the correct arrival time of a wavepacket at B. This ignorance is irrelevant—the collapse occurs without any intervention by the experimenter. And if the experimenter feels desperately unhappy about this ignorance of timing, she could go to the extra expense of installing an extra sensor or chopper in the beamline before the first grating in Fig. 3, to collect timing information for arriving atoms. But in our experimental procedure we do not need to determine the explicit value of the time of the "informative" non-detection, because our test is decided entirely and conclusively by the production of scattered light, not by detector-timing data of any kind.

[17] For the purposes of the proposed experimental test, we are not obligated to exploit the doubling of the pre-collapse time from 5 ns to 10 ns, and we could try to perform the experiment within the 5 ns interval. But the doubling of the pre-collapse time has obvious practical advantages.

[18] M. S. Chapman et al., Phys. Rev. Lett. **75**, 3783 (1995).

[19] A slight deviation from parallelism is tolerable, provided the change in beam-to-beam separation within the width of the focal spot of the laser (see below) is small compared with the wavelength of the laser light.

[20] We need wavefunction collapse at both spacetime points A and B, because collapse at only one point—say, by a measurement that detects the atom at A—would leave undefined the wavefunction in the interior of A's past light cone (the interior of this light cone would be blank, as in Fig. 1), so it would be impossible to calculate the radiation produced by the twin wavepackets at times that precede the time of A.

[21] Calculations of emission of dipole radiation in transitions between atomic states are given by M. L. Goldberger and K. M. Watson, *Collision Theory*, Wiley, New York (1964), p. 464, Eq. 177; p. 476, Eq. 229.

[22] In an earlier version of the proposal for such an experiment I suggested detection of the difference between coherent and incoherent emission of the fluorescent light by tuning the beam-to-beam distance to an odd number of laser half wavelengths, in the hope that this would actually cancel the emission of light entirely, because the transition matrix for excitation of the entire two-wavepacket atom from the ground state would be zero. However, this is not true, because when the distance between the two atomic wavepackets is large, the dipole approximation cannot be applied to the superposition of the two wavepackets, but must be applied separately to each wavepacket, and then the emitted radiated waves must be superposed; that is, the two wavepackets must be treated as two

separate dipole emitters radiating coherently, like two separate radio stations radiating coherently but with a phase difference. The radiation then does not cancel, but instead forms a complicated fringe pattern of closely-spaced interference beams. arXiv:1311.5840v2 corrects this mistake of arXiv:1311.5840v1.

[23] I assume that the measurements at A and B produce a total collapse, so the initial coherent superposition of left and right wavepackets collapses into a single wavepacket, left *or* right. This is called a "selective" measurement by Hellwig and Kraus, and it produces a well-defined collapsed wavefunction (a "pure" state), although with an overall random phase factor contributed by the macroscopic measuring apparatus, because of the decoherence of the apparatus states [for a discussion of decoherence in macroscopic systems, see R. Omnès, *Understanding Quantum Mechanics,* Princeton University Press, Princeton (1999), Chapters 17, 18]. Hellwig and Kraus also contemplate an alternative collapse involving a "nonselective" measurement, which does not apply to a pure state, but to a mixed state, described by a density matrix. Both of these alternative formulations of H-K collapse yield the same average scattering rate for laser photons but the "selective" formulation adds the (self-evident) prediction that if the atom is detected by, say, the left detector, then the scattered photon is emitted by the left beam.

[24] J. Schmiedmayer et al., "Optics and interferometry with atoms and molecules," in P. R. Berman, ed., *Atom Interferometry*, Academic Press, San Diego (1997), p. 17.

[25] Such a large beam separation is of course easily attainable in ordinary interferometers operating with light waves. This suggests a "complementary" experiment with coherent parallel light beams in which scattering is provoked (or not) by a transverse atom-laser beam crossing the light beams. Laser beams of bosonic atoms have been mentioned in the literature [see D. E. Pritchard et al., Ann. Phys. (Leipzig) **10**, 35 (2001)], but a 3-m long beam is not within practical reach.

[26] J. A. Wheeler and W. J. Zureck, eds., *Quantum Theory and Measurement*, Princeton University Press, Princeton (1983), p. 769.

[27] Both Perez, op. cit., and Omnès, op. cit., emphasize that random phases are more important for collapse than irreversible amplification. Instead of demanding an irreversible act of amplification, Bohr should have demanded an irreversible act of phase-randomnization, or "decoherence." Macroscopic systems acquire random phases extremely quickly (see Omnès, op. cit.); microscopic atomic wavepackets acquire random phases more slowly, but still sufficiently quickly so we can ignore the time span for this randomnization in our experiment. Time spans for randomnization can be calculated from the "master equation" for development of decoherence (see R. Omnès, op. cit., p. 200) or from collision rates of atoms and the associated random phase deviations accumulated by these collisions [see R. Omnès, *The Interpretation of Quantum Mechanics*, Princeton University Press, Princeton (1994), pp. 319-323; and E. Joos and H. D. Zeh, Condensed Matter, **59**, 223 (1985)].

The Joos and Zeh calculation is more complicated than it needs to be. Here is a simple estimate for decoherence of a wavepacket: For a collision with random impact parameters and random thermal velocities of the target atoms, the typical random change in kinetic energy of the incident atom is of order of magnitude of the kinetic energy itself (if the masses of both atoms are of the same order). The random phase deviation that the incident atom's wavepacket accumulates in the time $t$ between this collision and the next is then of the order of $K.E. \times t / \hbar$ (this phase deviation arises from the discrepancy between the phase and group velocities of the wavepacket; in contrast to a light wavepacket, the kinetic-energy deviation results in a phase deviation at the peak of the atom wavepacket that increases linearly with time). Even for an incident sodium atom with a low thermal velocity—say, 500 m/s—the accumulated phase deviation for the 0.0002-ns, 0.1-nm trip from one collision to the next in an ordinary solid or liquid environment is of the order of 10 radians, which means it takes only a few collisions to achieve complete phase randomnization of the wavepacket, which transforms the initial pure state of the wavepacket into a mixed state, described by a statistical probability distribution. The total time span required for this is only 0.001 ns or so, which is negligible on the scale of Fig. 4.

The final mixed state that the wavepacket reaches by interactions with the atoms of the detector can be calculated by scattering theory or by transport theory; see Goldberger and Watson, op. cit., Section 11.5. Something similar was done much earlier by Heisenberg [*The Physical Principles of the Quantum Theory,* University of Chicago (1930), Chapter V], but with no attention to alterations of the phase.

[28] Throughout my discussion of the experiment I focused on a backward collapse at the speed of light. However, speeds anywhere in the supra-infinite range (with a collapse hypersurface that lies between the past light cone and the conventional constant-time hypersurface) could also be explored. This merely requires shifting the laser beam upward in Fig. 3, to a position closer to the detectors.